\documentstyle[preprint,aps]{revtex}

\begin{document}
\draft
\title{Pseudogaps and Extrinsic Losses in Photoemission Experiments
on Poorly Conducting Solids}
\author{Robert Joynt}
\address{Department of Physics \\
University of Wisconsin-Madison \\
1150 University Avenue \\
Madison, WI 53706 \\}
\date{\today}
\maketitle

\begin{abstract}
A photoelectron, on being emitted from
a conducting solid, may suffer a substantial energy change
through ohmic losses that can drastically alter the lineshape
on the meV scale which is now
observable due to improved resolution.
Almost all of this energy loss takes place
after the electron leaves the solid.  These losses are expected to be important in
isotropic materials with relatively low conductivity, such as 
certain colossal magnetoresistance manganates and 
very electrically anisotropic materials, such as one-dimensional conductors.
Such effects may also be important in 
the interpretation of photemission in high-T$_c$
superconductors.  In all these materials, the electric
field of the photoelectron can penetrate the system.  In particular, 
extrinsic losses of this type can mimic pseudogap effects
and other peculiar features of photoemission in cubic manganates.  This
is illustrated with the case of La$_{0.67}$Ca$_{0.33}$MnO$_3$.
\end{abstract}
\newpage
%\pacs{PACS Nos. 95.30.Cq, 97.10.Cv, 97.60.Jd}

In the past few years, the resolution of photoemission (PE) experiments
has improved to the range of 10 meV or less, and this has allowed finer  
details of electronic structure to be observed, including 
the "pseudogap" - a depression of intensity at the chemical potential $\mu$.  
Pseudogaps have been observed in a wide variety of materials: 
quasi-one-dimensional (1D) systems, both inorganic (Ta Se$_4$)$_2$I 
\cite{dardel} and organic (TTF-TCNQ) \cite{zwick}, quasi-2D
systems such as the underdoped high-T$_c$ materials \cite{loeser}, and most recently
3D systems: 
the colossal magnetoresistance (CMR) manganates \cite{park},
\cite{dessau1}.
In many cases, interesting temperature 
dependences of these pseudogaps have been observed.  The origin of pseudogaps is 
among the most fundamental problems of present-day condensed matter physics.      
Because the most direct way to see them 
is with PE, it is well to understand this measurement very thoroughly.

A somewhat disturbing aspect of the current situation is that, although the 
resolution has greatly improved, isolated resolution-limited peaks are not 
the rule in angle-resolved photoemission (ARPES) 
data that detect pseudogaps.  There is a suggestion here that some 
extrinsic broadening mechanism is at work or that a large unexplained
background is present \cite{norman}.
   
The conventional wisdom interpretation of 
ARPES data is that at a given wavevector $\vec{k}$, the ideal intensity 
$I(\omega)$ is proportional to $A(\vec{k}, \omega)$, 
the spectral function for 
a single hole.  The observed intensity, at least near $\mu$, 
is broadened only because of the finite instrumental resolution.
$A(\vec{k}, \omega)$ is, in this context, an "intrinsic" 
quantity.  The outgoing electron either 
suffers a large energy loss due, for example, to plasmon emission or ionization, 
or suffers no loss.  In the former case, the electron is not detected or its energy is 
sufficiently far from threshold that it is ignored; in the latter case, the 
electron is detected and its measured distribution is a faithful reflection 
of the intrinsic distribution in the solid.

This conventional picture of the 
photoemission process is reconsidered in this report
for certain important classes of materials, namely 
those which are 'poor conductors'.  The working definition of this phrase
is a DC resistivity $\rho_0$ which exceeds the 
Mott value $\sim 100 \mu \Omega$-cm.  
I will argue that electrons emitted from such materials are subject to 
losses of the order of a few tens of meV after they leave the surface.  
At low resolutions, these processes are usually not important, but for high-resolution
experiments they cannot be ignored.

If an electron is emitted normally at speed $v$ 
from very near 
a clean surface and leaves the sample without undergoing 
significant energy 
loss, then
the Fourier transform of the electric field inside the 
material is
\begin{equation}
\vec{E}(\vec{r},\omega) = \frac{-e}{2 \pi v} 
\frac{2}{1+\epsilon(\omega)} \int_0^{\infty} d z'
e^{-i \omega z'/v} \frac{\vec{r}-z'\hat{z}}{|\vec{r}-z'\hat{z}|^3}.
\label{eq:elec}
\end{equation}   
The surface is the $x-y$ plane, $-e$ is the charge on the electron,
and $\epsilon(\omega)$ is 
the bulk dielectric function.
This electric field can set up currents in the bulk. 

To arrive at this expression certain approximations have been made.
The expression for $\vec{E}$ does not hold when the 
charge is within a few atomic layers of the surface: to model 
the short-time, high-frequency losses,
a proper treatment using the surface dielectric function would be required.
I do not attempt this here, as only the low-frequency loss is of 
interest.  I assume the normal skin effect - 
the wavevector dependence of $\epsilon(\omega)$ has been neglected.  
At high frequencies
or for very low temperatures for clean systems, the anomalous skin effect
should be taken into account.  
The factor $2/(1+\epsilon)$ in Eq.\ \ref{eq:elec} gives image charge 
and screening effects and proves critical.

These formulas depend on the assumption that the material is cubic.
The important special case of 
emission along the z-axis of a tetragonal material 
may be treated by the same method, and the image charge 
factor becomes $2/(1+\sqrt{\epsilon_{xx}\epsilon_{zz}})$.  Thus the 
absorption 
is {\it strongly enhanced} in a layered conducting material where we 
expect $|\epsilon_{xx}|>>|\epsilon_{zz}|$ at the relevant frequencies.  
Similar remarks apply to the orthorhombic 1D conductor case, 
but the calculations become far more complicated 
and no simple expression comparable to Eq.\ \ref{eq:elec} 
could be derived.
 
The currents set in motion by the field will produce 
ohmic loss.  These will be represented in the observed energy 
of the electron.  Classically, the
total energy loss is given by 
\begin{equation}
Q =  \frac{1}{2} \int_{-\infty}^{\infty} d \omega \int d^3r~
        \Re \sigma(\omega) |\vec{E}(\vec{r}, \omega)|^2
  =  \frac{2e^2}{\pi v} {\cal C} \int_0^{\infty} 
        d \omega \frac{L(\omega)}{\omega}
\end{equation}
where ${\cal C} \approx 2.57$
and $L(\omega) = \Re \sigma (\omega)/ |1+\epsilon(\omega)|^2$.

This classical calculation corresponds to a 
quantum-mechanical one.  In fact, as
constant electron velocity was assumed, it is the Born 
approximation.  Because the field is appropriately 
screened by the dielectric function, I term it the screened
Born approximation.  This approximation should be valid 
for those electrons 
whose energy loss is small compared with their 
total energy.  This ratio is
of order 50 mev / 20 eV $\sim$ 2.5 $\times 10^{-3}$ for experimental 
parameters of interest.  
The relative differential probability is obtained by setting
\begin{equation}
Q = \hbar \int_0^{\infty} \omega P(\omega) d \omega
\end{equation}
where $P(\omega)$ is the relative differential 
probability of losing energy $\hbar \omega$.  Hence
\begin{equation}
P(\omega) = \frac{2 e^2 {\cal C} L(\omega)}{\pi \hbar v \omega^2}
\end{equation}
This expression is general, and is of course related to well-known
formulas in electron-energy-loss spectroscopy \cite{x1}.  
Its relevance to PE has been noted before \cite{x2}.
Refs.\ \cite{x1} and \cite{x2} are concerned with
plasmon and other losses in the electron volt range.
Recent work on processes occurring when the electron is still
inside the material has also clarified the losses in this energy range
\cite{norman}, while highlighting the lack of explanation 
of background in the millielectron volt range.
 
Note at this point that $P(\omega)$ at low frequencies
is greater for systems with low conductivity.
Because $\epsilon(\omega) \sim 4 \pi i \sigma / \omega$ we
have $P(\omega) \sim \sigma / \omega^2 |\epsilon|^2 \sim 1/ \sigma $.

Quantum mechanics requires some probability for
forward scattering $P_0$, or that the electron
loses zero energy.  Thus, the total normalization is given by the equation
\begin{equation}
1 = P_0 + \int_0^{\infty} P(\omega) d \omega
\end{equation}  $P_0$ depends on an integration over all energies.  Because the 
dielectric function is usually not known quantitatively over the entire
range of energies, $P_0$ is difficult to evaluate.  For interpretation of data
it is best treated as a fit parameter.
  
I now apply these ideas to angle-integrated PE, saving
ARPES for later work.  I assume that $P(\omega)$
is independent of emission angle, which should be true for the
near-normal emissions typical for the incident photon energies
used in most cases.  The observed intensity $I(\omega, T)$, 
if electrons are emitted from a material
with a temperature-independent density of states
$N(\omega)$, is
\begin{equation}
I(\omega, T) = P_0(T) N(\omega) f(\omega) 
+ \int_0^{\infty} P(\omega-\omega', T) N(\omega') f(\omega') d \omega'
\label{eq:int}
\end{equation}
which must then be convoluted with an instrumental resolution
function.  The 'intrinsic' temperature dependence comes entirely from the
Fermi function $f(\omega)$, but this dependence is very minor; 
I restrict the argument to relatively low T.  

I first consider a model system 
for illustrative purposes.  For emission at a given $\vec{k}$
(ARPES), the observed intensity should consist of a main peak
at $\omega = \epsilon_{\vec{k}}$ and an asymmetric tail below this,
a rather common observation.  For the angle-integrated quantity 
$I(\omega, T)$, we obtain a two-component result according to Eq.\ \ref{eq:int}:
the actual density of states $N(\omega)$ and a downshifted loss
curve.  Can this mimic a pseudogap ?  Let $N(\omega) = N_0$ over some
wide energy range ($\sim eV$) below $\mu$, so that
there is no actual pseudogap.
Let the model system be a Drude conductor:
\begin{equation}
\sigma(\omega) = \frac{\sigma_0}{1 - i \omega \tau(T)}
\label{eq:drude}
\end{equation}
This expression is then substituted in Eq.\ \ref{eq:int} to produce Fig.\ 1
for two conductivities.  The parameters for the dashed curve 
are: $\rho_0 = 1/ \sigma_0 = 110~ \mu \Omega-$cm
and $\tau = 4 \times 10^{-14}$ s.  The parameters for the solid curve 
are: $\rho_0 = 1/ \sigma_0 = 44.5~ m \Omega$-cm
and $\tau = 10^{-16}$ s.  Both curves have $P_0 = 0.01$ and 
T = 38 K (k$_B$ T = 3.3 meV).  $\sigma_0 / \tau$ is held 
fixed in the figure.  The point is very simple: in the Drude model
$\sigma_0/ \tau$ is just $ n e^2 /m^*$, where $n$ is the carrier 
concentration and $m^*$
is the effective mass, so that all of the 
temperature dependence in the conductivity 
occurs in the relaxation time, as in a conventional
metal with no gap or pseudogap.  
The changes in the observed intensity arise
entirely from extrinsic effects.  The other 
parameters are held fixed as well.
Note that these DC resistances are very high by the standards of 
ordinary metal physics, but quite typical
of the CMR systems at temperatures comparable to or 
below the metal-insulator
(M-I) transition.  In the highly resistive state, the
fields penetrate into the material and losses are high, whereas the loss
is relatively low for the high conductivity state that screens the field.
The plots have been
normalized in the conventional manner by setting the intensities equal at
a binding energy where they have leveled out - here at $-350$ meV.  The results are
not very sensitive to this number.
The popular midpoint 
method used to determine a "pseudogap"
would give a value of about 50 meV.      
The dashed curve represents a system at the Mott conductivity - 
the borderline at which 
the loss effects become important.  In 
good metals with $\rho_0 < 100~ \mu \Omega$-cm
losses become negligible, and the observed spectra 
reflect the actual density of states faithfully.  

The curves demonstrate that in a system with a M-I transition,
the observed intensity will change due to "extrinsic" effects.  In 
general, there will be a motion of weight away from the Fermi energy
as one approaches the insulating state.  If such motion is observed in
experiment it may not have anything to do with an actual
pseudogap in the density of states.

Considering now actual spectra, angle-integrated PE on the CMR 
material La$_{0.67}$Ca$_{0.33}$MnO$_3$
shows a number of unusual and striking features, represented by the points
in Fig.\ 2, taken from Park {\it et al.} \cite{park}.  The material has a
M-I transition at 260 K.  
In the metallic state at 80 K, there is a strong 
negative slope in $I(\omega)$ for at least 0.6 eV below $\mu$.  
There is a sharp break in slope at $\mu$, 
presumably indicative of a nonzero
density of states at $\mu$.  In the insulating state at 280 K
there appears to be no
Fermi edge at all - the observed intensity is flat at $\mu$
and weight has moved back from $\mu$.  There is even upward 
curvature in the data, as opposed to the downward curvature of
the Fermi function.  
There is nothing in the usual theory of metals to account 
for any of these
observations, and they certainly do not agree with 
band calculations \cite{pickett}. 
These features have been taken to indicate a 
pseudogap \cite {dessau2}, but they can be produced by extrinsic effects.

In Fig.\ 2, I plot the data points at two temperatures against the theory
(Eq.\ \ref{eq:int}).  It is necessary to take a model for the 
frequency-dependent conductivity, which is not entirely Drude-like
in the manganates.  I have adopted a simplified version of
the model of 
Okimoto {\it el al.} \cite{okimoto} in which there
is a frequency-independent part $\sigma_{01}$ 
and a Drude part $\sigma_d(\omega)$ 
which is as in Eq.\ \ref{eq:drude}.  This introduces an additional parameter
$r= \sigma_{01}/\sigma_d(0)$ which measures the relative strength of the
two components.  The authors of Ref.\ \cite{okimoto}
base their model on the analysis of 
their data on optical conductivity of La$_{1-x}$Sr$_x$MnO$_3$ which is 
isoelectronic to the the calcium-doped system.  Some sharp structure in the
optical data, presumably due to phonons, is neglected in 
the model.  If included, it might account
for some of the additional small structure observed in PE.

The parameters for the upper curve are:
$\tau =5 \times 10^{-14}$s, 
$\rho_0 = 1 / \sigma_d(0) = 0.296~ m\Omega-$cm, 
T= 80 K, $P_0 = 0.0025$ and $r = 0.2$.
The parameters for the lower curve are:
$\tau =10^{-14}$ s, 
$\rho_0 = 1 / \sigma_d(0) = 1.48~ m \Omega$-cm, 
T= 280 K, $P_0 = 0$ and $r = 0.25$.  The curves are normalized to agree
at a binding energy of 600 meV.

Again in this case, there is no change in the underlying density of states
and the change in the theoretical intensity is entirely
due to extrinsic effects.

To make a convincing case for a pseudogap from PE
material a careful analysis of data using Eq.\ \ref{eq:int}
is required to extract a pseudogap from experimental data.
This suggests that meaningful investigation of electronic structure
in poorly conducting materials requires a combination of 
PE with optical conductivity and electron energy loss measurements.
This allows us to apply Eq.\ \ref{eq:int} and back out the density of
states.  A simple check can always be made.  The inelastic part
of the spectrum is inversely proportional to the 
speed of the outgoing electron, as may be seen from Eq.\ \ref{eq:elec}.  
Hence, to be genuine, a 
pseudogap must be present in the observed intensity at all incoming
photon energies.
 
One may make some qualitative statements about 
the current situation in some of the more important classes of 
materials beyond the CMR manganates.

In good-quality high-T$_c$ superconductors, 
the conductivity in the $a-b$ plane
typically exceeds the Mott value.  However,  
the conductivity along the $c-$axis
is often less.  Thus, these materials form a marginal case for the loss mechanism
described here.  There are other very strong indications that the 
pseudogap in the underdoped materials is quite real.  
ARPES itself shows that the pseudogap is momentum-dependent, which the 
loss is not.  There is also corroboration from other experiments,
tunneling being perhaps the most persuasive because it is also a direct 
measure of the DOS \cite{renner}.  On the other hand, details of 
lineshapes may still be affected by extrinsic processes in high-T$_c$ materials.  
A distinct sharpening of
quasi-particle-like peaks is often observed as the 
temperature is lowered and the DC conductiviity increases, suggesting a decrease
in energy loss.    

In 1D systems, conductivity in two directions is very low,
and one might expect the losses to be substantial.  Intriguingly, it often appears
to be the case that the gap or pseudogap measured in PE is greater than that
given by other experiments.  In (TaSe$_4$)$_2$I, for example, the PE gap
at low temperatures is about 500 meV, whereas other experiments 
give values near 250 meV \cite{nic}.
Another well-known example is TTF-TCNQ.  At room temperature, 
DC transport data may
be interpreted as that of a highly anisotropic gapless 
metal \cite{cooper}, but
a pseudogap of 120 meV is observed in ARPES \cite{zwick}.  
These are only two of numerous examples of this puzzling mismatch which 
can be cited in 1D conductors.
Such results are a strong indication that extrinsic processes are 
influencing the photoelectron spectrum in these systems.

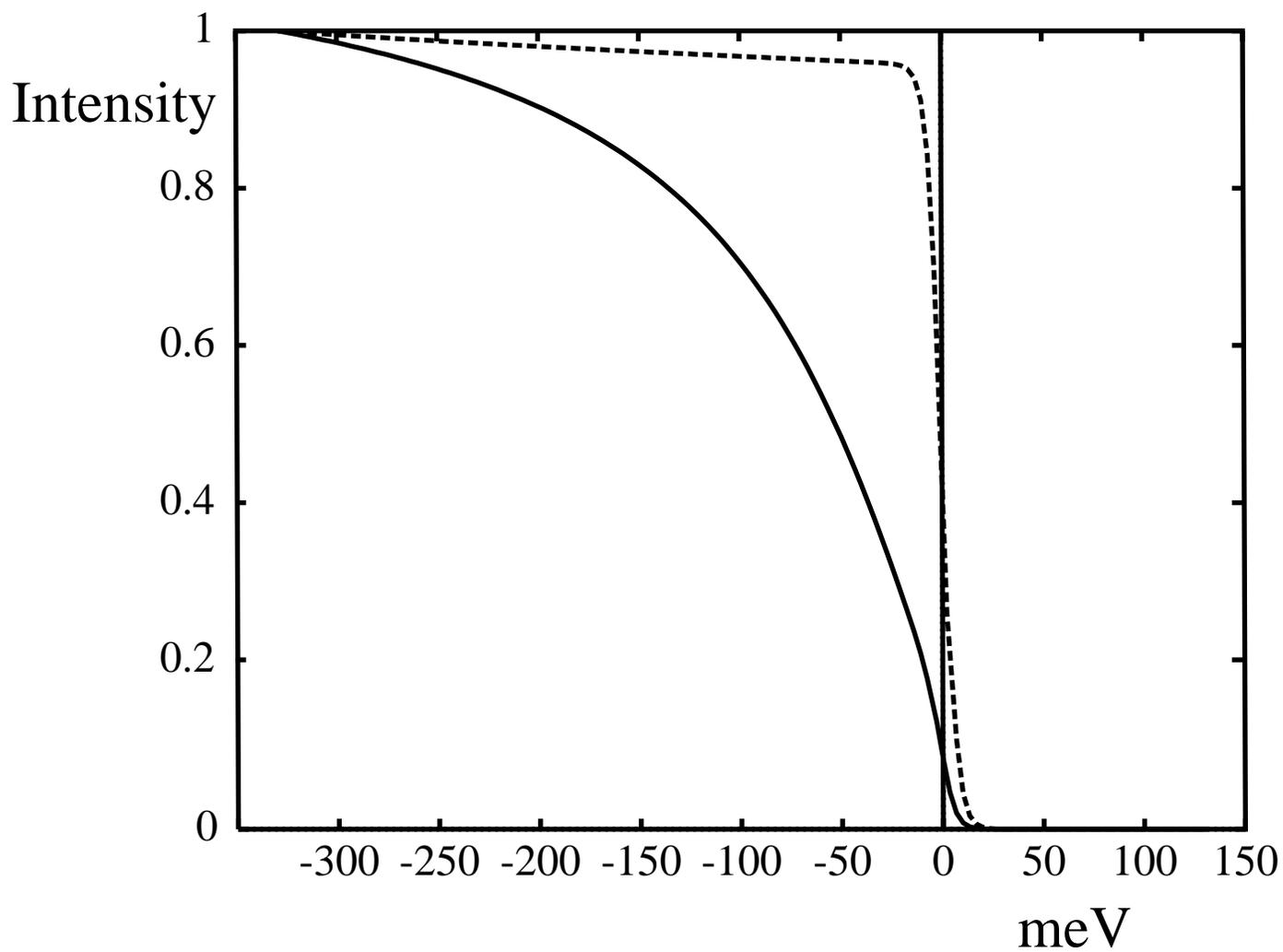
\begin{figure}
\caption[]{Calculated photoemission spectra from an idealized solid with 
a constant density of states.  The curves differ only in the 
relaxation times: the dashed curve has $\tau = 10^{-14}$ s whereas the 
solid curve has $\tau = 4 \times 10^{-16}$ s.  The emitted electrons 
thereby lose different amounts of energy on exiting the solid.}
\label{fig:illus}
\end{figure}

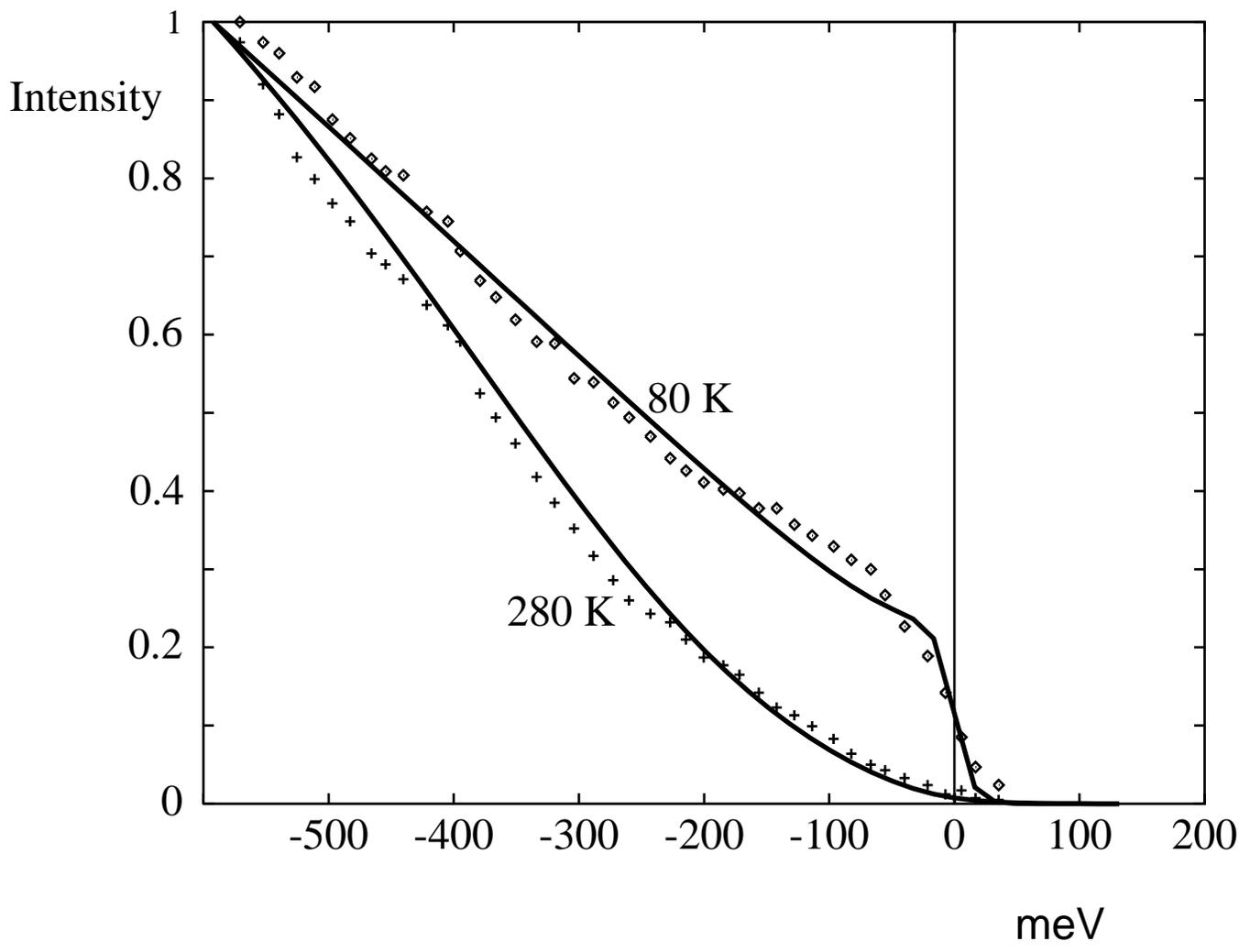
\begin{figure}
\caption[]{Calculated (lines) and observed (points)
photoemission spectra for La$_{0.67}$Ca$_{0.33}$MnO$_3$.  Parameters of the
fit are given in the text.  The lineshapes, which do not resemble 
ordinary densities of states, are strongly affected by inelastic 
processes, which indeed dominate the lower curve.  Points are taken from
Ref.\ [4].}
\label{fig:lcmo}
\end{figure}

\end{document}